\documentclass[pre,showpacs,twocolumn]{revtex4}

\usepackage{amsfonts}
\usepackage{amsmath}
\usepackage{amssymb}
\usepackage{graphicx}

\def\kbar{\protect\@kbar}
\def\@kbar{\relax \bgroup
\def\@tempa{\hbox{\raise.73\ht0
\hbox to0pt{\kern.25\wd0\vrule width.5\wd0 height.1pt
depth.1pt\hss}\box0}}\mathchoice{\setbox0\hbox{$\displaystyle
k$}\@tempa}{\setbox0\hbox{$\textstyle
k$}\@tempa}{\setbox0\hbox{$\scriptstyle
k$}\@tempa}{\setbox0\hbox{$\scriptscriptstyle k$}\@tempa}\egroup}

\begin{document}

\title{\textbf{Weak-Chaos Ratchet Accelerator}}
\author{Itzhack Dana and Vladislav B. Roitberg}
\affiliation{Minerva Center and Department of Physics,\\
Bar-Ilan University, Ramat-Gan 52900, Israel}

\begin{abstract}
Classical Hamiltonian systems with a mixed phase space and some asymmetry
may exhibit chaotic ratchet effects. The most significant such effect is a
directed momentum current or acceleration. In known model systems,
this effect may arise only for sufficiently strong chaos. In this paper, a
Hamiltonian ratchet accelerator is introduced, featuring a momentum current
for \emph{arbitrarily weak chaos}. The system is a realistic generalized kicked rotor and is exactly solvable to some extent, leading to analytical expressions for the momentum current. While this current arises also for relatively strong chaos, the \emph{maximal} current is shown to occur, at least in one case, precisely in a limit of arbitrarily weak chaos.\newline
\end{abstract}

\maketitle

\begin{center}
\textbf{I. INTRODUCTION}\\[0pt]
\end{center}

Classical \cite{hr1,hr2,hr3,hr4,hr5,hr6} and quantum 
\cite{hr3,hr6,qr1,qr2,qr3,qr4,qr5,qr6} Hamiltonian ratchets have attracted a
considerable theoretical interest during the last decade. Also, several
kinds of quantum ratchets have been experimentally realized using
atom-optics methods with cold atoms or Bose-Einstein condensates 
\cite{qr7,qr8,qr9,qr10}. The classical Hamiltonian ratchet effect is a directed
current in the chaotic region generated by an unbiased force (having zero
mean in space and/or time) and due to some spatial and/or temporal asymmetry 
\cite{hr1,hr2,hr3,hr4,hr5,hr6}. This is analogous to the ordinary ratchet
effect \cite{rat,ral}, but with deterministic chaos replacing the usual noisy
environment. Dissipation, an important ingredient in ordinary ratchets for
breaking time-inversion symmetry, is absent in Hamiltonian ratchets.

A well studied class of systems are those described by time-periodic
Hamiltonians $H(x,p,t)$ for which both the force $F=-\partial H/\partial x$
and the velocity $v=\partial H/\partial p$ are periodic in $x$ and $F$ has zero mean over $(x,t)$. The classical ratchet current is usually defined
as the average of $v$ over $(x,p,t)$, where $(x,p)$ is restricted to the
chaotic region, see, e.g., Refs. \cite{hr2,hr3}. It is assumed that $v$ is
bounded, e.g., by Kolmogorov-Arnol'd-Moser (KAM) tori. Then, necessary
conditions for the ratchet current to be nonzero are the breaking of some symmetry
and a \emph{mixed} phase space featuring ``transporting" stability islands which propagate in the $x$ direction \cite{hr3}. In the presence of bounding KAM tori, 
one can get, in principle, a well-defined ratchet current also in
near-integrable regimes, corresponding to relatively \emph{weak} and 
\emph{local} chaos.

A different and much more significant Hamiltonian ratchet effect was
discovered in work \cite{hr6} for generalized kicked-rotor systems
satisfying the well-known KAM scenario. Namely, for sufficiently strong
kicking, there exist no KAM tori bounding the chaotic motion in the momentum
($p$) direction, and one then gets \emph{strong global} chaos. In addition,
there may arise transporting \textquotedblleft accelerator-mode\textquotedblright\ islands \cite{ai} propagating in the $p$ direction. This can lead, under some asymmetry conditions, to a \emph{\textquotedblleft ratchet acceleration\textquotedblright }, i.e., a nonzero mean \emph{momentum} velocity (rather than the usual position velocity $v$) of the global chaotic region \cite{hr6} (see more details in Sec. II).
Quantum analogs of the classical ratchet acceleration were found in several
systems \cite{qr3,qr4,qr5,qr6}, either for special, \textquotedblleft
quantum-resonance\textquotedblright\ values of a scaled Planck constant 
$\hbar $ \cite{qr3,qr4,qr5} or for generic values of $\hbar $ \cite{qr6}.
Quantum-resonance ratchet accelerators have been experimentally realized in
recent works \cite{qr8,qr9}.

In this paper, we show for the first time that the phenomenon of ratchet
acceleration is \emph{not} limited to strong-chaos regimes. We introduce a
realistic Hamiltonian system exhibiting this phenomenon most significantly
in near-integrable regimes, corresponding now to \emph{arbitrarily weak} but 
\emph{global} chaos. The system is a generalized kicked rotor whose force
function has zero mean and is characterized by two nonintegrability
parameters $b_{1}$ and $b_{2}$. A global chaotic region in the $p$
direction arises also for arbitrarily small values of these parameters, i.e., 
the KAM scenario is not satisfied. As $b_{1},\ b_{2}\rightarrow 0$, this \textquotedblleft non-KAM\textquotedblright\ 
\cite{nkam} system tends to the well-known elliptic sawtooth map \cite{pa,gmz},
which has been used as a paradigmatic model of \textquotedblleft
pseudochaos\textquotedblright\ (dynamical complexity with zero Lyapunov
exponent) \cite{gmz,fv} in studies of both classical \cite{gmz,id} and quantum 
\cite{qesm} systems. We show that accelerator-mode islands exist for arbitrarily small $b_{1}$ and $b_{2}$. Then, when the system is asymmetric ($b_{1}\neq b_{2}$), a
ratchet acceleration $A$ may arise for arbitrarily weak chaos. In one particular case, we derive analytical expressions for $A$ as a function of $b_{1}$ and $b_{2}$.
Paths of maximal $A$ in the $(b_{1},b_{2})$ parameter space are determined. We then show that, in sharp contrast with the systems considered in work \cite{hr6}, $A$ is most significant for relatively small Lyapunov exponent and that its \emph{maximal} value is attained precisely in a limit $b_{1},\ b_{2}\rightarrow 0$ of arbitrarily weak chaos.

This paper is organized as follows. In Sec. II, we give a short background
on Hamiltonian ratchet accelerators. In Sec. III, we introduce our general model
system and describe its basic properties. In particular, in Sec. IIIC we derive the existence conditions for the main accelerator-mode islands of the system. In Sec. IV, analytical expressions for the ratchet acceleration $A$ in one case are obtained for all values of the parameters. In Sec. V, we show that the maximal value of $A$ is
attained in a limit of arbitrarily weak chaos. Conclusions are presented in Sec. VI. Detailed derivations of several analytical results are given in the Appendices.

\begin{center}
\textbf{II. BACKGROUND ON HAMILTONIAN RATCHET ACCELERATORS}\\[0pt]
\end{center}

The concept of Hamiltonian ratchet accelerator was introduced in Ref. 
\cite{hr6} by adaptation of a formalism developed in Refs. \cite{hr2,hr3}. We
give here a self-contained summary of these works, leading to the main
result (\ref{sum}) below, a sum rule for the ratchet acceleration $A$. We
shall focus on realistic models, the generalized kicked-rotor systems with
scaled Hamiltonian $H=p^{2}/2+KV(x,t)\sum_{s=-\infty }^{\infty }\delta (t-s)$, 
where $K$ is the nonintegrability parameter and the potential $V(x,t)$ is
periodic in $x$, $V(x+1,t)=V(x,t)$. Particular cases of these systems where
considered in Ref. \cite{hr6}. The map for $H$ from $t=s-0$ to $t=s+1-0$ is
given by
\begin{equation}
M:\ p_{s+1}=p_{s}+Kf_{s}(x_{s}),\ x_{s+1}=x_{s}+p_{s+1}\ \mathrm{mod}(1),   \label{M}
\end{equation}
where the force function $f_{s}(x)=-dV(x,t=s)/dx$. Because of the
periodicity of $V(x,t)$ in $x$, $f_{s}(x)$ satisfies the ratchet
(zero-flux) condition: $\langle f_{s}(x)\rangle =\int\limits_{0}^{1}f_{s}(x)dx\,=0$. As in Ref. \cite{hr6},  we shall assume that the kicking parameter $K$ is large enough that all
the rotational (\textquotedblleft horizontal\textquotedblright ) KAM tori
are broken. Then, there are no barriers to motion in the $p$ direction,
leading to a global and strongly chaotic region. These barriers cannot exist
if there are \textquotedblleft accelerator modes\textquotedblright , i.e.,
orbits that are periodic under map (\ref{M}) in the following sense: 
\begin{equation}
p_{s+m}=p_{s}+w,\ \ \ x_{s+m}=x_{s},  \label{am}
\end{equation}
where $m$ is the period and $w$, the winding number, is an nonzero integer. 
Periodic orbits can be defined in the generalized way (\ref{am}) ($w\neq 0$) 
due to the obvious
periodicity of the map (\ref{M}) in $p$ with period $1$. If an accelerator
mode is linearly stable, each of its points $(x_{s},p_{s})$ will be usually
surrounded by an island $\mathcal{I}_{s}$, an \textquotedblleft
accelerator-mode island\textquotedblright\ (AI). Because of (\ref{am}), 
$\mathcal{I}_{s+m}$ is just $\mathcal{I}_{s}$ translated by $w$\ in the $p$
direction. For arbitrary initial conditions $\mathbf{z}_{0}=(x_{0},p_{0})$
in phase space, the mean acceleration (momentum current/velocity) in $n$ iterations of (\ref{M}) is 
\begin{equation}
A_{n}(\mathbf{z}_{0})=\frac{p_{n}-p_{0}}{n}  \label{An}
\end{equation}
and the average of (\ref{An}) in some region $\mathcal{R}$ with area $S_{\mathcal{R}}$ is 
\begin{equation}
\langle A_{n}\rangle _{\mathcal{R}}=\frac{1}{S_{\mathcal{R}}}\int\limits_{
\mathcal{R}}\,A_{n}(\mathbf{z}_{0})d\mathbf{z}_{0},  \label{AEn}
\end{equation}
In the case that $\mathcal{R}$ is an AI $\mathcal{I}$ with winding number $w$, 
it follows from Eqs. (\ref{am})-(\ref{AEn}) that 
\begin{equation}
\lim_{n\rightarrow \infty }\langle A_{n}\rangle _{\mathcal{I}}=\nu =\frac{w}{
m}.  \label{AIn}
\end{equation}

Now, because of the periodicity of the map (\ref{M}) in $p$ with period $1$,
one can take also $p_{s+1}$ modulo $1$ in (\ref{M}), leading to a map $\bar{M
}$ on the unit torus $\mathbb{T}^{2}:\ 0\leq x,\ p<1$; this is the unit cell
of periodicity of map (\ref{M}). The reduced phase space $\mathbb{T}^{2}$
can be fully partitioned into the global chaotic region $\mathcal{C}$
with area $S_{\mathcal{C}}$ and all the stability islands $\mathcal{I}^{(j)}$ 
with areas $S_{j}$, where $j$ labels the island: 
$S_{\mathcal{C}}+\sum_{j}S_{j}=1$. The average acceleration (\ref{AIn}) of 
$\mathcal{I}^{(j)}$ is $\nu _{j}=w_{j}/m_{j}$, where $\nu _{j}=0$ for a
normal (non-accelerating) island. We then have the following sum rule
relating $\nu _{j}$ to the ratchet acceleration $A=\langle A\rangle _{
\mathcal{C}}=\lim_{n\rightarrow \infty }\langle A_{n}\rangle _{\mathcal{C}}$
of the global chaotic region: 
\begin{equation}
S_{\mathcal{C}}A+\sum_{j}S_{j}\,\nu _{j}=0.  \label{sum}
\end{equation}
Eq. (\ref{sum}) is easily derived from the obvious relation $\langle
A_{n}\rangle _{\mathbb{T}^{2}}=S_{\mathcal{C}}\langle A_{n}\rangle _{
\mathcal{C}}+\sum_{j}S_{j}\,\langle A_{n}\rangle _{\mathcal{I}^{(j)}}$ by
taking $n\rightarrow \infty $ and using $\langle A_{n}\rangle _{\mathbb{T}
^{2}}=0$, a result following straightforwardly from the map (\ref{M}): 
\begin{eqnarray}
\langle A_{n}\rangle _{\mathbb{T}^{2}} &=& \left\langle \sum\limits_{s=1}^{n}
\frac{p_{s}-p_{s-1}}{n}\right\rangle _{\mathbb{T}^{2}}=\frac{K}{n}
\sum\limits_{s=0}^{n-1}\int\limits_{\mathbb{T}^{2}}d\mathbf{z}
_{0}f_{s}(x_{s}) \nonumber \\
 &=& \frac{K}{n}\sum\limits_{s=0}^{n-1}\int\limits_{\mathbb{T}
^{2}}d\mathbf{z}_{s}f_{s}(x_{s})=0,  \label{proof}
\end{eqnarray}
where we used area preservation ($d\mathbf{z}_{0}=d\mathbf{z}_{s}$), the
invariance of $\mathbb{T}^{2}$ under $\bar{M}$, and the ratchet condition 
$\langle f_{s}(x)\rangle =0$. An immediate consequence of relation 
(\ref{sum}) is that $A$ vanishes if the map (\ref{M}) is
invariant under inversion, $(x,p)\rightarrow (-x,-p)$, i.e., one has the
inversion (anti)symmetry $f_{s}(-x)=-f_{s}(x)$. This is because under this
symmetry for each AI with mean acceleration $\nu _{j}\neq 0$ there exists an
AI with the same area but with mean acceleration $-\nu _{j}$. As we shall
see in the next sections for a simple case of $f_{s}(x)$, $A$ is generally
nonzero when AIs are present and inversion symmetry is absent.

\begin{center}
\textbf{III. THE GENERAL MODEL SYSTEM AND ITS BASIC PROPERTIES}\\[0pt]
\end{center}
\begin{center}
\textbf{A. General}\\[0pt]
\end{center}

The general model system introduced and studied in this paper is the
generalized kicked-rotor system described by a simple map (\ref{M}):
\begin{equation}
M:\ p_{s+1}=p_{s}+Kf(x_{s}),\ x_{s+1}=x_{s}+p_{s+1}\ \mathrm{mod}(1), 
\label{cpem}
\end{equation}
where $0<K<4$ and, for $0\leq x<1$, 
\begin{equation}
f(x)=
\begin{cases}
l_{1}x & \text{for $\ 0\leq x\leq b_{1}$}, \\ 
c-x & \text{for $b_{1}<x<1-b_{2}$}, \\ 
l_{2}(x-1) & \text{for $1-b_{2}\leq x<1,$}
\end{cases}
\label{fx}
\end{equation}
with $f(x+1)=f(x)$. Here $b_{1}$ and $b_{2}$ are positive parameters
with $b_{1}+b_{2}<1$ while $l_{1}$, $l_{2}$, and $c$, also positive, are
fixed by requiring $f(x)$ to be continuous and to satisfy the ratchet
condition $\int\limits_{0}^{1}f(x)dx\,=0$ (see Appendix A):
\begin{eqnarray}
l_{1} &=&\frac{(1-b_{1})(1-b_{1}-b_{2})}{b_{1}(2-b_{1}-b_{2})},\ \ l_{2}=
\frac{(1-b_{2})(1-b_{1}-b_{2})}{b_{2}(2-b_{1}-b_{2})}, \nonumber
\\
c &=&\frac{1-b_{2}}{2-b_{1}-b_{2}}.  \label{mmc}
\end{eqnarray}
Kicked systems with a smooth piecewise linear force function such as 
(\ref{fx}) have been studied either on the phase plane \cite{jk} or on a
cylindrical phase space \cite{plsm}, corresponding to the very special case of
map (\ref{cpem}) with $b_{1}=b_{2}=1/4$. Apparently, however, these systems
have not been considered yet in the context of Hamiltonian ratchet
transport, i.e., for general values of $b_{1}$ and $b_{2}$ with $b_{1}\neq
b_{2}$, leading to an asymmetric force function (\ref{fx}). This general
system is realistic since it may be experimentally realized using, e.g.,
optical analogs as proposed in Ref. \cite{jk}. As we shall see below, the
system generally does not satisfy the KAM scenario assumed in Sec. II, i.e.,
it is a non-KAM system.

\begin{center}
\textbf{B. Phase space and limit cases}\\[0pt]
\end{center}

The phase space of the map (\ref{cpem}) in the basic periodicity torus 
$\mathbb{T}^{2}$ ($0\leq x,\ p<1$) is illustrated in Fig. 1 for some values
of the parameters. 
\begin{figure}[th]
\centering\includegraphics[width=4.2cm,height=10.5cm]{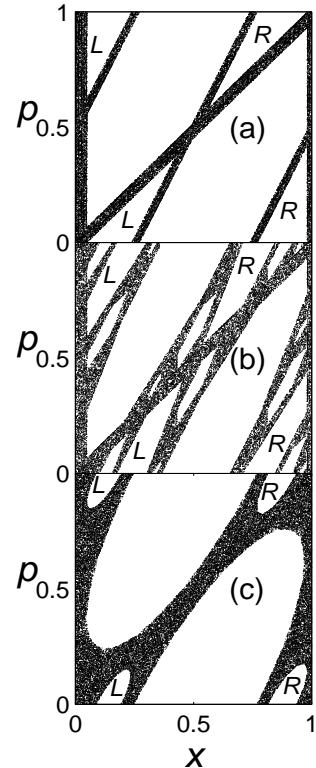}
\caption{Global chaotic regions of the map (\protect\ref{cpem}) within the
unit torus of periodicity $0\leq x,\ p<1$ for $b_{1}=0.05$, $b_{2}=0.02$,
and three values of $K$ corresponding to the following values of $\protect
\alpha $ in Eq. (\protect\ref{eig}): (a) $\protect\alpha =2\protect\pi /3$ 
($K=3$); (b) $\protect\alpha =6\protect\pi /5$ ($K\approx 3.618$); (c) 
$\protect\alpha =\protect\pi (\protect\sqrt{5}-1)/2$ ($K\approx 2.7247$).
These values of $\protect\alpha $ represent the three main cases discussed
in the text. The \textquotedblleft left\textquotedblright\ ($L$) and
\textquotedblleft right\textquotedblright\ ($R$) period-$1$ accelerator-mode
islands (AIs, see Sec. IIIC) are indicated in each case.}
\end{figure}
We clearly see in all cases a connected chaotic region encircling $\mathbb{T}
^{2}$ in both the $x$ and $p$ directions, implying global chaos and the
nonexistence of KAM tori bounding $p$. An understanding of this numerical
observation will be achieved here and in Sec. IIIC. We first consider here
the map (\ref{cpem}) in the limit of $b_{1},\ b_{2}\rightarrow 0$. From Eqs.
(\ref{mmc}) one has $l_{1}b_{1},\ l_{2}b_{2},\ c\rightarrow 1/2$ in this
limit, so that the function (\ref{fx}) tends to the sawtooth 
\begin{equation}
f(x)=1/2-x\ \ \mathrm{(}0\leq x<1\mathrm{)},\ \ \ f(x+1)=f(x),  \label{saw}
\end{equation}
with discontinuity at $x=0$. The map (\ref{cpem}) with (\ref{saw}) and 
$0<K<4 $ is the well-known elliptic sawtooth map (ESM) \cite{pa,gmz,id,qesm}
having the property that its linearization $DM$ is a constant $2\times 2$
matrix with eigenvalues $\lambda _{\pm }$ on the unit circle:
\begin{equation}
\lambda _{\pm }=\exp (\pm i\alpha ),\text{ \ }2\cos (\alpha )=2-K.
\label{eig}
\end{equation}
This means that orbits of the ESM which do \emph{not} cross the
discontinuity line $x=0$ lie on ellipses with average rotation angle 
$\alpha$. In general, however, an orbit will cross the $x=0$ line. Then, the
combination of the mod($1$) operation in (\ref{cpem}) with the local
ellipticity of the ESM will usually lead to a complex dynamics with zero
Lyapunov exponent, known as \textquotedblleft pseudochaos\textquotedblright\ 
\cite{gmz}. The phase space generally consists of the pseudochaotic region,
associated with all iterates of the discontinuity line \cite{pa}, and a set
of islands. More specifically, one has to distinguish between three main
cases of the ESM, illustrated in Fig. 2 for the same values of $K$\ as in
Fig. 1: (a) The \textquotedblleft integrable\textquotedblright\ case of
integer $K=1,\ 2,\ 3$ [corresponding to $\alpha /(2\pi )=1/6,\ 1/4,\ 1/3$ in
(\ref{eig})]; in this case, no pseudochaos arises and the phase space
consists just of a finite number of \textquotedblleft
separatrix\textquotedblright\ lines (iterates of the discontinuity line)
bounding a finite number of islands, see Fig. 2(a). (b) The case of
non-integer $K$\ with rational $\alpha /(2\pi )$ in (\ref{eig}); in this
case, numerical work \cite{pa} indicates that one has an infinite set of
islands and that the pseudochaotic region is a fractal with zero area, see
Fig. 2(b) and exact results for the fractal dimension of such regions
in other maps with discontinuities \cite{fv}. (c) The case of irrational 
$\alpha /(2\pi )$; here one typically has again an infinite set of islands
but the pseudochaotic region appears numerically to cover a finite area 
\cite{pa}, see Fig. 2(c). Since the momentum $p$ assumes all values on the
discontinuity line and is thus unbounded, the pseudochaos [or the separatrix
in case (a)] is global. 
\begin{figure}[th]
\centering\includegraphics[width=4.2cm,height=10.5cm]{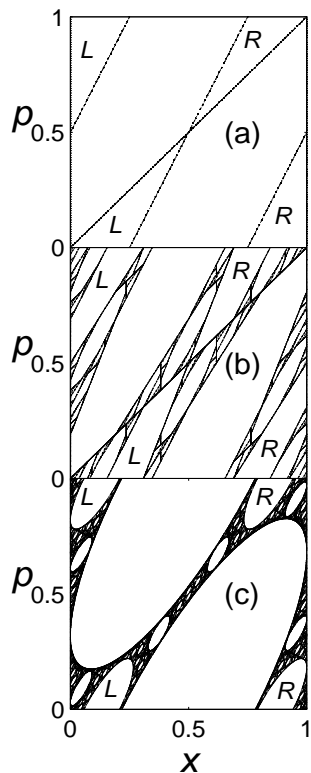}
\caption{Global \textquotedblleft pseudochaotic\textquotedblright\ regions
of the ESM for the same values of $\protect\alpha $ (or $K$) as in Fig. 1.
In practice, these regions were generated by iterating a large initial
ensemble using the map (\protect\ref{cpem}) with very small 
$b_{1}=b_{2}=10^{-7}$. The initial ensemble covered uniformly
the vertical \textquotedblleft hyperbolic\textquotedblright\ strip
(\ref{B}). The left and right AIs are again indicated.}
\end{figure}

For finite and small $b_{1}$ and $b_{2}$, the continuous map (\ref{cpem})
may be considered as a perturbed ESM, with the discontinuity line replaced
by a vertical strip $B$ of width $b_{1}+b_{2}$ in $\mathbb{T}^{2}$ (see also caption of Fig. 2):
\begin{equation}\label{B}
B:\ \ 0\leq p\leq 1\ \ \mathrm{and}\ \ 0\leq x\leq b_{1}\ \ \mathrm{or}\ \
1-b_{2}\leq x<1.  
\end{equation}
This should be contrasted with the perturbed ESM in Ref. \cite{id} for which the discontinuity line is not removed by the perturbation. The linearization $DM$ of (\ref{cpem}) is again a constant $2\times 2$ matrix in each of of the three intervals in Eq. (\ref{fx}). In the middle interval, it is the same matrix as for the ESM, 
with stability eigenvalues (\ref{eig}). In the other two intervals, where the
strip (\ref{B}) is located, $DM$ can be easily shown to have real positive
eigenvalues $\lambda _{\pm }$ with, say, $\lambda _{+}>1$ and $\lambda
_{-}=\lambda _{+}^{-1}<1 $, i.e., there is local hyperbolicity. One can then expect that already for small $b_{1}$ and $b_{2}$ a global chaotic region with positive
Lyapunov exponent will emerge from the vertical strip (\ref{B}) and will
replace the global pseudochaos (or separatrix) for $b_{1}=b_{2}=0$. This can
be clearly seen by comparing Figs. 1 and 2. The nature of the chaotic region
will be further discussed in the next sections, where it will be shown
numerically that the Lyapunov exponent indeed tends to zero as $b_{1},\
b_{2}\rightarrow 0$.

\begin{center}
\textbf{C. Accelerator-mode islands (AIs) and their existence conditions}\\[0pt]
\end{center}

We show here that AIs for the map (\ref{cpem}) rigorously exist in broad ranges of the parameters, including arbitrarily small values of $b_{1}$ and $b_{2}$. This
exactly implies global and arbitrarily weak chaos. We shall consider only 
period-$1$ AIs, associated with stable accelerator modes satisfying Eq. (\ref{am})
with $m=1$ and $w\neq 0$. As we shall see, there appear to be no
higher-period AIs at least in the case of $K=3$ on which we shall focus
from next section on. The initial conditions $(x_{0},p_{0})$ for $m=1$
stable periodic orbits in Eq. (\ref{am}) must necessarily lie in the middle
interval in Eq. (\ref{fx}), $b_{1}<x_{0}<1-b_{2}$, since only in this interval the matrix $DM$ exhibits stability eigenvalues (\ref{eig}). Then, from Eqs. (\ref{am}), (\ref{cpem}), and 
(\ref{fx}), we get:
\begin{equation}
x_{0}=c-\frac{w}{K},\ \ \ p_{0}=0\ \text{\textrm{mod}}(1),  \label{xp}
\end{equation}
\begin{equation}
b_{1}<c-\frac{w}{K}<1-b_{2}.  \label{cam}
\end{equation}
For $w=0$ one has a non-accelerating stable fixed point $(x_{0}=c,p_{0}=0)$, the center of a normal (non-accelerating) island. Let us show that $w$ may take only two nonzero values and this only in some interval of $K$: 
\begin{equation}
w=\pm 1,\ \ 2\leq K<4.  \label{wK}
\end{equation}
In fact, from Eqs. (\ref{fx}) and (\ref{mmc}) it follows that the maximal value of $|f(x)|$ is $\max (l_{1}b_{1},l_{2}b_{2})<1/2$. Then, since $w=Kf(x_{0})$ from Eqs. (\ref{am}) and (\ref{cpem}), we have $\left\vert w\right\vert \leq \lbrack K/2]$, where $[\ ]$ denotes integer part. This implies, for $0<K<4$, that $w$ may take
the only nonzero values of $\pm 1$ provided $2\leq K<4$.

Now, according to Eq. (\ref{xp}) for $x_{0}$, the values of $w=1$ and $w=-1$
should correspond, respectively, to a \textquotedblleft left\textquotedblright\ 
($L$) and \textquotedblleft right\textquotedblright\ ($R$) AI, see Figs. 1 and
2. An explicit existence condition for the left AI ($w=1$) is derived, after some simple algebra, from the left inequality in (\ref{cam}) using Eq. (\ref{mmc}) for $c$: 
\begin{equation}
b_{2}<F(b_{1})\equiv \frac{Kb_{1}^{2}+(1-2K)b_{1}+K-2}{K-1-Kb_{1}},
\label{lai}
\end{equation}
see also note \cite{note}. It is easily verified that the right inequality in (\ref{cam}) is identically satisfied. Similarly, the existence condition for the right AI is $b_{1}<F(b_{2})$. One thus has three cases (compare with Fig. 3 for $K=3$): 
\begin{figure}[th]
\centering\includegraphics[width=6.5cm,height=6.5cm]{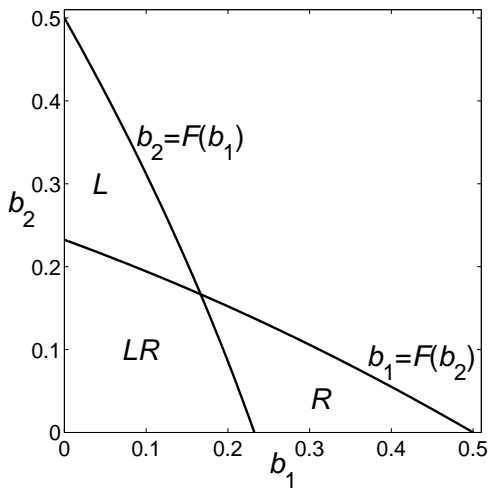}
\caption{The curves $b_{2}=F(b_{1})$ and $b_{1}=F(b_{2})$ [with $F(b_{1})$ given by Eq. (\protect\ref{lai}) for $K=3$], defining the domains of existence of the left and right period-$1$ AI for $K=3$ in the $(b_{1},b_{2})$ plane. In
domain $L$ ($R$), only the left (right) AI exists. In domain $LR$, both AIs
exist. No AIs exist elsewhere.}
\end{figure}

(a) Both AIs $L$ and $R$ exist (see, e.g., Figs. 1 and 2) if
\begin{equation}
b_{2}<F(b_{1})\text{ }\mathrm{and}\text{ }b_{1}<F(b_{2}).  \label{ca}
\end{equation}
Clearly, this will be always satisfied for $K>2$ and sufficiently small
$b_{1}$ and $b_{2}$ since $F(b_{1})\approx (K-2)/(K-1)$ for $b_{1}\ll 1$ in
Eq. (\ref{lai}); thus, both AIs exist in the arbitrarily weak chaos regime.
For $K=3$, this case corresponds to the domain $LR$ in Fig. 3.

(b) Only one AI, say the right one $R$, exists (as, e.g., in Fig. 4) if
\begin{equation}
b_{2}\geq F(b_{1})\text{ }\mathrm{and}\text{ }b_{1}<F(b_{2}).  \label{cb}
\end{equation}
\begin{figure}[th]
\centering\includegraphics[width=6.5cm,height=6.5cm]{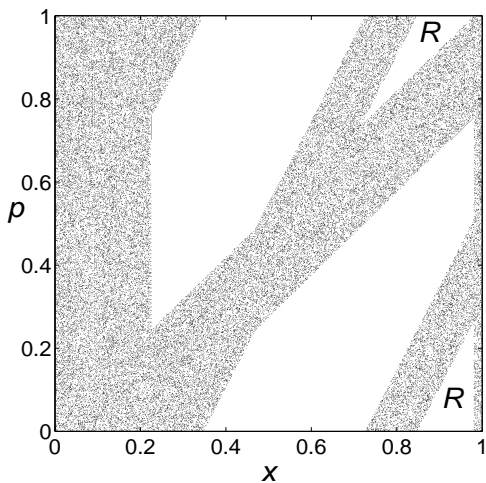}
\caption{Global chaotic region for $K=3$, $b_{1}=0.226$, and $b_{2}=0.02$.
For these values of $b_{1}$ and $b_{2}$ only the right period-$1$ AI exists,
compare with Fig. 3.}
\end{figure}
Similarly, if $b_{2}<F(b_{1})$ and $b_{1}\geq F(b_{2})$ only the left AI $L$
exists. For $K=3$, this case corresponds to the domain $R$ or $L$ in Fig. 3.

(c) No AIs exist if
\begin{equation}
b_{2}\geq F(b_{1})\text{ }\mathrm{and}\text{ }b_{1}\geq F(b_{2}).  \label{cc}
\end{equation}

It is easy to show from the expression for $F(b_{1})$ in Eq. (\ref{lai}) (see also note \cite{note}) that $F(b_{1})\leq (K-2)/(K-1)-b_{1}$. One then gets from Eqs. (\ref{ca})-(\ref{cc}) a simple necessary condition for the existence of at least one AI:
\begin{equation}
b_{1}+b_{2} < \frac{K-2}{K-1}.  \label{ccs}
\end{equation}

For the symmetric system ($b_{1}=b_{2}$), the condition (\ref{lai}) reads 
$b_{1}<F(b_{1})$, which can be significantly simplified:
\begin{equation}
b_{1}<\frac{1}{2}-\frac{1}{K}.  \label{lais}
\end{equation}
It follows from condition (\ref{lais}) that no period-$1$ AIs can exist if 
$b_{1}=b_{2}\geq 1/4$, for any value of $K$ in the relevant interval of 
$2\leq K<4$. This is consistent with the known fact that bounding KAM tori
exist for some $K$ if $b_{1}=b_{2}=1/4$ \cite{plsm}, which is apparently the
only case of map (\ref{cpem}) studied until now.

\begin{center}
\textbf{IV. RATCHET ACCELERATION FOR }$\boldsymbol{K=3}$\\[0pt]
\end{center}

In this section, the ratchet acceleration $A$ in the case of $K=3$ will be
calculated analytically in the framework of a plausible assumption (see below),
supported by extensive numerical evidence and exact results. To use the sum
rule (\ref{sum}), we first identify the global chaotic region $\mathcal{C}$
in the basic periodicity torus $\mathbb{T}^{2}$. Let us denote by $C$ the set of all iterates of the vertical strip $B$ in Eq. (\ref{B}) under $\bar{M}$, i.e., the map (\ref{cpem}) modulo $\mathbb{T}^{2}$ (see Sec. II):
\begin{equation}
C=\bigcup_{s=-\infty }^{\infty }\bar{M}^{s}B.  \label{C}
\end{equation}
Exact results for the set (\ref{C}) are derived in Appendices B-E. Here we note that orbits which never visit $B$ (and thus also $C$) are all stable since they lie entirely within the middle interval in (\ref{fx}) where the linearized map $DM$ has stability eigenvalues (\ref{eig}). Thus, the global chaotic region $\mathcal{C}$ must
be entirely contained within $C$, in agreement with our expectation at the end of Sec. IIIB.  Our extensive numerical studies indicate that $C$ is indistinguishable from $\mathcal{C}$, compare, e.g., Figs. 1(a) and 4 with Figs. 11 and 12 in Appendix C. In fact, finite-time Lyapunov exponents of orbits starting from initial conditions covering $B$ uniformly were found to be all strictly positive. We shall therefore assume in what follows that $\mathcal{C}$ precisely coincides with $C$. The rest of phase space outside $\mathcal{C}$ consists of no more than three stability regions [see, e.g., Figs. 1(a) and 4]: The left AI $L$ ($w=1$), the right AI $R$ ($w=-1$), and a normal island ($w=0$) lying between $L$ and $R$. Using the sum rule 
(\ref{sum}) with $\nu _{j}=w_{j}$ [since $m=1$ in Eq. (\ref{AIn})], we then get a formula for the ratchet acceleration: 
\begin{equation}
A=\frac{S_{R}-S_{L}}{S_{C}}.  \label{Ae}
\end{equation}
Exact expressions for the areas $S_{L}$, $S_{R}$, and $S_{C}$ are derived in
Appendices D and E using simple geometry; see Eqs. (\ref{SL}), (\ref{SR}),
and (\ref{SC1})-(\ref{SC3}) there. Inserting these expressions in formula
(\ref{Ae}), we obtain after some algebra explicit results for $A$ in
different cases:

(a) If both AIs exist, i.e., case (\ref{ca}), 
\begin{equation}
A=\frac{(b_{1}-b_{2})[1-3(b_{1}+b_{2})]}{2(2-b_{1}-b_{2})(b_{1}+b_{2})}.
\label{Aa}
\end{equation}

(b) If only one AI, say the right one, exists, i.e., case (\ref{cb}),
\begin{equation}
A=\frac{(2-3b_{2}-3c)^{2}}{
6(b_{1}+b_{2})-6(b_{1}+b_{2})^{2}+(3c-3b_{1}-1)^{2}}.  \label{Ab}
\end{equation}

(c) If no AIs exist, i.e., case (\ref{cc}), $A=0$, of course.

In general, the results (\ref{Aa}) and (\ref{Ab}) were found to agree very
well with numerical calculations of $A$, see examples at the end of the next section.
This is additional evidence for the validity of the basic assumption above 
concerning the chaotic region, $\mathcal{C}=C$.
%\newpage

\begin{center}
\textbf{V. MAXIMAL RATCHET ACCELERATION FOR
ARBITRARILY WEAK CHAOS}\\[0pt]
\end{center}

In this section, we show that the maximal ratchet acceleration $A$ for $K=3$
is attained in a limit $b_{1},\ b_{2}\rightarrow 0$ of arbitrarily weak chaos. 
In Fig. 5, we plot $|A|$ as function of $b_{1}$ and $b_{2}$ using formulas
(\ref{Aa}), (\ref{Ab}), and $A(b_2,b_1)=-A(b_1,b_2)$. 
\begin{figure}[ht]
%\centering
\hspace*{-1.4cm}\includegraphics[width=8.5cm,height=8.5cm]{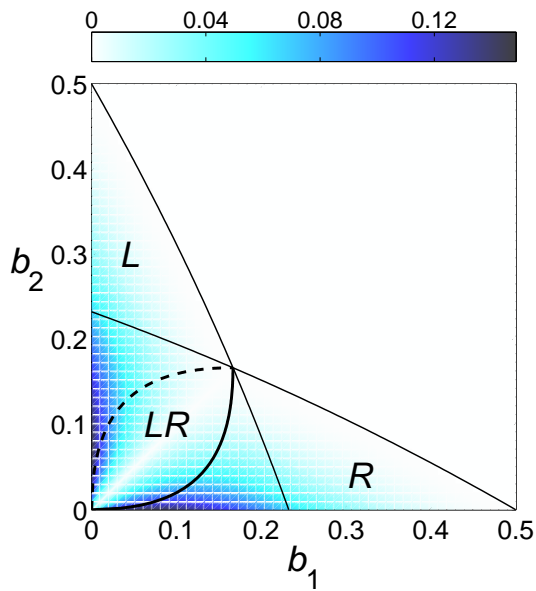}
\caption{(Color online) Pseudocolor plot of $|A|$ as function of $b_{1}$ and $b_{2}$ for $K=3$. The thin solid lines (defining the three domains $L$, $R$, and $LR $) are the same as those in Fig. 3. The thick solid line in domain $LR$
is the \textquotedblleft maximal\textquotedblright\ path (\protect\ref{b2max})
on which $A$ is given by Eq. (\protect\ref{Amax}). The dashed line is the
maximal path $b_{2}(b_{1})$ [defined similarly to (\protect\ref{b2max})] on
which $A<0$.}
\end{figure}
The Lyapunov exponent $\sigma $ of the chaotic region as function of $b_{1}$
and $b_{2}$ was calculated numerically with high accuracy and is plotted in
Fig. 6. 
\begin{figure}[ht]
%\centering
\hspace*{-1.4cm}\includegraphics[width=8.5cm,height=8.5cm]{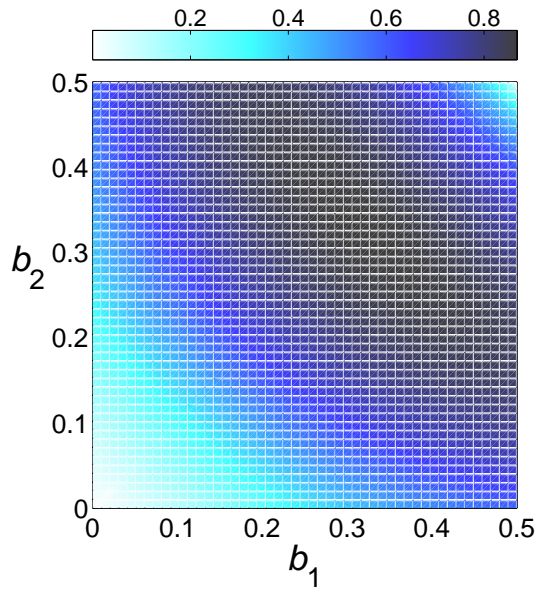}
\caption{(Color online) Pseudocolor plot of the Lyapunov exponent $\protect\sigma $ of the chaotic region as function of $b_{1}$ and $b_{2}$ for $K=3$.}
\end{figure}
As one could expect, $\sigma $ vanishes in the limit of $b_{1},\
b_{2}\rightarrow 0$, where the map (\ref{cpem}) tends to the ESM (see Sec.
IIIB). It is clear from Fig. 5 that $|A|$ assumes its largest values
in the parameter domain $LR$, where both AIs exist. We shall therefore focus
on this domain in which $A$ is given by Eq. (\ref{Aa}). We shall first
calculate analytically the value of $b_{1}$ where $\left\vert
A(b_{1},b_{2})\right\vert $ is maximal at fixed $b_{2}$; this will define a
path $b_{1}(b_{2})$ in the $(b_{1},b_{2})$ plane [a path $b_{2}(b_{1})$ can
be similarly defined]. We then show that $\left\vert
A(b_{1},b_{2})\right\vert $ is maximal on this path in the limit of $b_{1},\
b_{2}\rightarrow 0$. Let us take the partial derivative of the function 
(\ref{Aa}) with respect to $b_{1}$ and require that $\partial A/\partial b_{1}=0$. After a tedious but straightforward calculation, we find that the latter equation reduces to a quadratic one with the only positive root: 
\begin{equation}
b_{1}(b_{2})=\frac{2\left[ 5b_{2}(1-b_{2})\right] ^{1/2}-b_{2}(7-6b_{2})}{
5-6b_{2}}.  \label{b2max}
\end{equation}
The path (\ref{b2max}) corresponds to the lower curve in Fig. 5, with 
$b_{1}\geq b_{2}$. This curve starts at $b_{1}=b_{2}=0$, with $b_{1}\approx 2
\sqrt{b_{2}/5}$ for $b_{2}\ll 1$, and terminates at $b_{1}=b_{2}=1/6$, 
on the boundary of the $LR$ domain. For $b_{2}\leq 1/6$, we find that 
$\partial ^{2}A/\partial b_{1}^{2}<0$ at the value (\ref{b2max}) of $b_{1}$,
which thus corresponds to a local maximum. From Eqs. (\ref{Aa}) and 
(\ref{b2max}), the ratchet acceleration on the path (\ref{b2max}) is: 
\begin{equation}
A(b_{2})=\frac{\left\{ 5-6\left[ 5b_{2}(1-b_{2})\right] ^{1/2}\right\} ^{2}}{
20\left\{ 5-4b_{2}-\left[ 5b_{2}(1-b_{2})\right] ^{1/2}\right\} }.
\label{Amax}
\end{equation}
In the limit of $b_{2}\rightarrow 0$ ($b_{1}\approx 2\sqrt{b_{2}/5}$),
we get from Eq. (\ref{Amax}):
\begin{equation}
\lim_{b_{2}\rightarrow 0}A(b_{2})=1/4.  \label{Am}
\end{equation}
After a simple but lengthy calculation, we find that the function 
(\ref{Amax}) satisfies $\partial A/\partial b_{2}<0$ for $b_{2}\leq 1/6$. Thus, 
$A(b_{2})$ decreases monotonically from $1/4$ (at $b_{1}=b_{2}=0$) to $0$ (at $b_{1}=b_{2}=1/6$) on the path (\ref{b2max}). Since this path gives the single extremum (a local maximum) of $A(b_{1},b_{2})$ for $b_{1}\geq b_{2}$ at fixed $b_{2}$
and since $A(b_{1},b_{2})=0$ for $b_{1}=b_{2}$, we conclude that in the
lower part ($b_{1}\geq b_{2}$) of the $LR$ domain $A(b_{1},b_{2})\geq 0$ and 
$A(b_{1},b_{2})$ assumes its maximal value of $1/4$ in the limit $b_{1},\ b_{2}\rightarrow 0$ of arbitrarily weak chaos on the path (\ref{b2max}).

Since $A(b_{2},b_{1})=-A(b_{1},b_{2})$, in the upper 
($b_{2}>b_{1}$) part of the $LR$ domain $A(b_{1},b_{2})<0$ and
$A(b_{1},b_{2}) $ assumes its maximal negative value of $-1/4$ in the limit
of $b_{1},\ b_{2}\rightarrow 0$ on a path $b_{2}(b_{1})$ (the upper curve in
Fig. 5), defined similarly to $b_{1}(b_{2})$. The difference between the
limiting values of $A(b_{1},b_{2})$ on the two paths reflects the
discontinuity of the ESM, i.e., the map (\ref{cpem}) in the limit of
$b_{1},\ b_{2}\rightarrow 0$. In general, $\left\vert
A(b_{1},b_{2})\right\vert $ can assume in this limit all values $<1/4$ on
other, \textquotedblleft non-maximal\textquotedblright\ paths. For example,
on the straight-line path $b_{1}=b_{2}/a$, where $a$ is some arbitrary
constant, we find from Eq. (\ref{Aa}) that 
\begin{equation}
\lim_{b_{2}\rightarrow 0}A(b_{2})=\frac{(1-a)}{4(1+a)}.  \label{Ama}
\end{equation}
We remark that the path (\ref{b2max}) is tangent to the $b_{1}$ axis at 
$b_{1}=b_{2}=0$, since $b_{1}\approx 2\sqrt{b_{2}/5}$ for $b_{2}\ll 1$. 
Similarly, the second maximal path $b_{2}(b_{1})$ is tangent to the $b_{2}$
axis in this limit. Thus, as expected, the maximal value of $\left\vert
A\right\vert =1/4$ is associated with the largest possible \textquotedblleft
asymmetry\textquotedblright , $b_{1}/b_{2}=\infty $ or $b_{2}/b_{1}=\infty $
[$a=0$ or $a=\infty $ in Eq. (\ref{Ama})].

Figs. 7 and 8 show plots of $A$ versus the Lyapunov exponent $\sigma $ for small $b_{2}$ on both the maximal path (\ref{b2max}) and the path $b_{1}=3b_{2}$. We see in both plots an excellent agreement between the values of $A$ calculated numerically and those calculated from formulas (\ref{Aa}), (\ref{Ab}), and (\ref{Amax}).
\begin{figure}[th]
\centering\includegraphics[width=6.5cm,height=6cm]{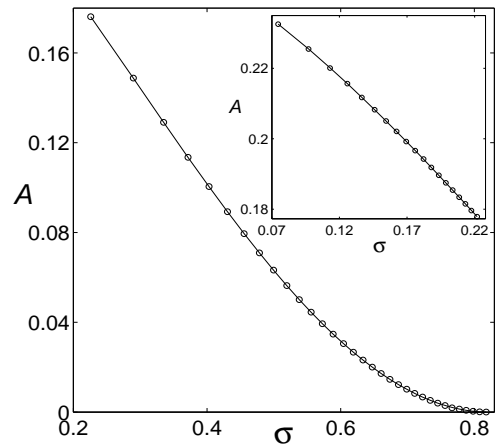}
\caption{Circles: Numerical results for $A$ versus the Lyapunov
exponent $\protect\sigma$ on the maximal path (\protect\ref{b2max}) with $
b_2 $ distributed uniformly on the interval $0.005\leq b_2\leq 0.165$; these
results were obtained by averaging $(p_n-p_0)/n$ ($n=120000$) over an
ensemble of $10^4$ initial conditions $(x_0,p_0)$ in the
chaotic region, i.e., all having positive finite-time Lyapunov exponents. 
Solid line: The analytical result (\protect\ref{Amax}). The
inset shows the continuation of the main plot to smaller values of $b_2$,
distributed uniformly on the interval $0.00025\leq b_2\leq 0.00475$. In this
interval of very weak chaos, $A$ is close to its maximal value $1/4$, see 
Eq. (\protect\ref{Am}).}
\end{figure}
\begin{figure}[th]
\centering\includegraphics[width=6.5cm,height=6cm]{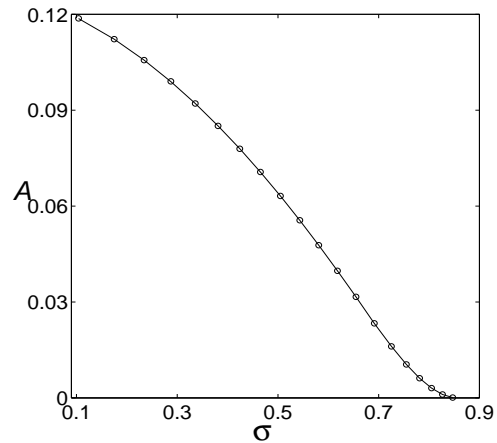}
\caption{Circles: Numerical results for $A$ versus $\protect\sigma$ on
the path $b_{1}=3b_{2}$ with $b_2$ distributed uniformly on the interval 
$0.005\leq b_2\leq 0.1$ (this path crosses both domains $LR$ and $R$, see
Fig. 3); these results were obtained as described in the caption of Fig. 7.
Solid line: Analytical results from formulas (\protect\ref{Aa}) and (\protect
\ref{Ab}). For small $\sigma$, $A\approx 0.12$, close to the predicted limiting value of $A=0.125$ from Eq. (\ref{Ama}) ($a=1/3$).}
\end{figure}

\begin{center}
\textbf{VI. CONCLUSIONS}\\[0pt]
\end{center}

In this paper, we have introduced a realistic non-KAM system exhibiting, in weak-chaos regimes, the most significant Hamiltonian ratchet effect of directed acceleration. The system, defined by the generalized standard map (\ref{cpem}) with (\ref{fx}), may be viewed as a perturbed elliptic sawtooth map (ESM) with a perturbation that removes the ESM discontinuity. Then, the global weak chaos featured by the system may be generally considered as a perturbed global pseudochaos. Our main exact result is that for $K=3$ the maximal ratchet acceleration $A$ is attained precisely in a limit $b_{1},\ b_{2}\rightarrow 0$ of arbitrarily weak chaos with infinite asymmetry parameter ($b_{1}/b_{2}=\infty $ or $b_{2}/b_{1}=\infty $). Despite this fact, the limiting system is interestingly the \emph{completely symmetric} ESM (see phase spaces in Fig. 2). By continuity considerations, one expects that at least for values of $K$ sufficiently close to $K=3$ one should again observe a significant increase of the absolute value $\vert A\vert$ of the acceleration as the chaos strength decreases. We have verified this numerically in parameter regimes where good accuracy could be achieved within the limitations of our available computational resources. An example is shown in Fig. 9.

\begin{figure}[th]
\centering\includegraphics[width=6.5cm,height=6cm]{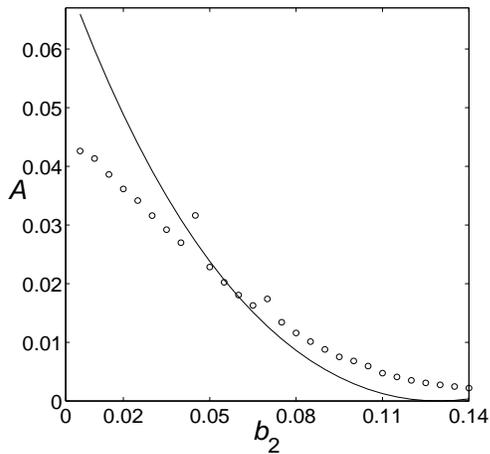}
\caption{Circles: Numerical values of $A$ (obtained as described in the
caption of Fig. 7) for $K=2.99$ on the path $b_{1}=0.25$ and $0.005\leq
b_{2}\leq 0.14$; this path lies entirely in domain $R$, see Fig. 3 for the
nearby value of $K=3$. The Lyapunov exponent on the path varies in the
interval $0.4787<\protect\sigma <0.8326$. Solid line: Analytical results
from formula (\protect\ref{Ab}) for $K=3$.}
\end{figure}
Our main result that the strongest Hamiltonian ratchet effect can arise in a limit of arbitrarily weak chaos has apparently no analog in ordinary ratchets if chaos is viewed as the deterministic counterpart of random noise. In fact, a sufficiently high level of noise is essential for the functioning of ordinary ratchets or Brownian motors \cite{rat,ral}. Actually, it was recently shown that for a L\'{e}vy ratchet the current decreases algebraically with the noise level \cite{ral}, in clear contrast with our results.

The quantized version of our non-KAM system may be experimentally realized using, e.g., optical analogs as proposed in Ref. \cite{jk} and is expected to exhibit in general a rich variety of quantum phenomena, including the quantum signatures of the weak-chaos ratchet acceleration. The study of these phenomena is planned to be the subject of future works.
 
\begin{center}
\textbf{ACKNOWLEDGMENTS}\\[0pt]
\end{center}

This work was partially supported by BIU Grant No. 2046.
\newline

\begin{center}
\textbf{APPENDIX A}\\[0pt]
\end{center}

We derive here Eqs. (\ref{mmc}). First, continuity of the function 
(\ref{fx}) at $x=b_{1}$ and $x=1-b_{2}$ implies that
\begin{equation}
l_{1}b_{1}=c-b_{1},\ \ \ l_{2}b_{2}=1-b_{2}-c.  \label{lb}
\end{equation}
Then, using Eqs. (\ref{fx}) and (\ref{lb}) in the ratchet condition 
$\int_{0}^{1}f(x)dx\,=0$, we find that
\begin{equation*}
\int_{0}^{1}f(x)dx=c-\frac{b_{1}}{2}c+\frac{b_{2}}{2}(1-c)-\frac{1}{2}=0,
\end{equation*}
yielding the expression for $c$ in Eqs. (\ref{mmc}). After inserting this
expression in Eqs. (\ref{lb}), we get the expressions for $l_{1}$ and $l_{2}$
in Eqs. (\ref{mmc}).

\begin{center}
\textbf{APPENDIX B: THE REGION} $\boldsymbol{C}$ \textbf{FOR} 
$\boldsymbol{K=3}$\\[0pt]
\end{center}

In this Appendix and in the next ones, we derive, for $K=3$, exact results
for the region $C$ in Eq. (\ref{C}). As mentioned in Sec. IV, several arguments and extensive numerical evidence indicate that $C$ coincides with the chaotic region $\mathcal{C}$ for $K=3$. We show here that one has the simple relation 
\begin{equation}
C=C^{\prime }=B\cup \bar{M}B\cup \bar{M}^{2}B.  \label{Cs}
\end{equation}
To show this, we first denote
\begin{equation}
\bar{B}^{(1)}=\bar{M}B-B\cap \bar{M}B,  \label{Bb1}
\end{equation}
\begin{equation}
\bar{B}^{(2)}=\bar{M}\bar{B}^{(1)}-B\cap \bar{M}\bar{B}^{(1)}.  \label{Bb2}
\end{equation}
We derive below the relation
\begin{equation}
\bar{M}\bar{B}^{(2)}\subseteq B.  \label{mb2}
\end{equation}
Then, from the definition of $C^{\prime }$ in Eq. (\ref{Cs}) and from Eqs. 
(\ref{Bb1})-(\ref{mb2}) it follows that
\begin{equation}
\bar{M}C^{\prime }\subseteq C^{\prime }.  \label{mc}
\end{equation}
Eq. (\ref{mc}) and the fact that $\bar{M}$ is area preserving imply that 
$\bar{M}C^{\prime }=C^{\prime }$ or $\bar{M}^{-1}C^{\prime }=C^{\prime }$.
Thus, $C^{\prime }=\bar{M}^{s}B\cup \bar{M}^{s+1}B\cup \bar{M}^{s+2}B$ for
all integers $s$, which is possible only if $C^{\prime }$ is equal to $C$ in
Eq. (\ref{C}). Relation (\ref{Cs}) is thus proven.

To derive Eq. (\ref{mb2}), we start by obtaining an explicit expression for 
$\bar{B}^{(1)}$ in Eq. (\ref{Bb1}). For $K=3$, the iterate of any initial condition 
$(x_{0},p_{0})$ under $\bar{M}$ satisfies $p_{1}=x_{1}-x_{0}\ \mathrm{mod}(1)$,
$x_{1}=x_{0}+p_{0}+3f(x_{0})\ \mathrm{mod}(1)$. Clearly, when $p_{0}$
varies in $[0,1)$ at fixed $x_{0}$, $x_{1}$ varies in the whole interval 
$[0,1)$. Then, taking $(x_{0},p_{0})$ in $B$ and using Eqs. (\ref{B}) and 
(\ref{Bb1}), we get
\begin{equation}
\bar{M}B=\left\{ (x,p)|0\leq x<1,\ x-b_{1}\leq p\leq x+b_{2}\right\} 
\mathrm{mod}(\mathbb{T}^{2}),  \label{B1}
\end{equation}
\begin{equation}
\bar{B}^{(1)}=\left\{ (x,p)|b_{1}<x<1-b_{2},\ x-b_{1}\leq p\leq
x+b_{2}\right\}.  \label{B1e}
\end{equation}
The region (\ref{B1e}) is a strip (parallelogram) of slope $1$, shown in
Fig. 10(b) and corresponding to the strip $B$ in Fig. 10(a).
\begin{figure}[th]
\centering\includegraphics[width=4.2cm,height=10.5cm]{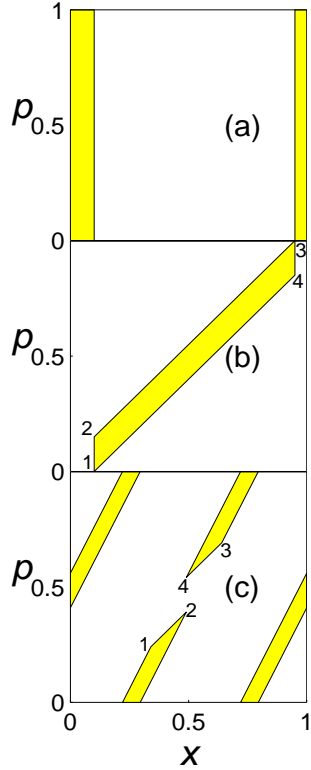}
\caption{(Color online) The figure shows, for $K=3$, $b_{1}=0.1$, and 
$b_{2}=0.05$: (a) the strip $B$, see Eq. (\protect\ref{B}); (b) the region 
$\bar{B}^{(1)}$, see Eq. (\protect\ref{Bb1}) or (\protect\ref{B1e}); (c) the region 
$B^{(2)}$, see Eq. (\protect\ref{B2}) or (\protect\ref{B2e}).}
\end{figure}
Next, we determine the region 
\begin{equation}
B^{(2)}=\bar{M}\bar{B}^{(1)}.  \label{B2}
\end{equation}
The second iterate $(x_{2},p_{2})$ of $(x_{0},p_{0})$ under $\bar{M}$, with 
$(x_{0},p_{0})\in B$ and $(x_{1},p_{1})\in \bar{B}^{(1)}$, is given by
\begin{equation}
p_{2}=p_{1}-3(x_{1}-c)\ \mathrm{mod}(1)=-2x_{1}-x_{0}+3c\ \mathrm{mod}(1),
\label{p2}
\end{equation}
\begin{equation}
x_{2}=x_{1}+p_{2}\ \mathrm{mod}(1)=-2x_{1}+p_{1}+3c\ \mathrm{mod}(1).
\label{x2}
\end{equation}
From Eq. (\ref{p2}) and the first Eq. (\ref{x2}), we find that
\begin{equation}
p_{2}=2x_{2}+x_{0}-3c\ \mathrm{mod}(1).  \label{p2e}
\end{equation}
Eq. (\ref{p2e}) and the second Eq. (\ref{x2}) imply that the region (\ref{B2}) 
is the following set of phase-space points: 
\begin{eqnarray}
& &B^{(2)}=\left\{ (x,p)\right\} : \nonumber \\
& &\begin{cases}
x=-2x_1+p_1+3c\ \mathrm{mod}(1),\ \ (x_1,p_1)\in B^{(1)}, \\ 
p=2x+x_0-3c\ \mathrm{mod}(1),\ \ -b_2\leq x_0\leq b_1.
\end{cases}
\label{B2e}
\end{eqnarray}
The region (\ref{B2e}) is clearly a parallelogram of slope $2$ folded into
$\mathbb{T}^{2}$, as shown in Fig. 10(c). The region $\bar{B}^{(2)}$
in Eq. (\ref{Bb2}) is given by Eq. (\ref{B2e}) with $x$ restricted to the
interval $b_{1}<x<1-b_{2}$. Then, using also Eq. (\ref{p2e}), we get for $(x_{0},p_{0})\in B$ and $(x_{2},p_{2})\in \bar{B}^{(2)}$:
\begin{eqnarray}
& &\bar{M}\bar{B}^{(2)}=\left\{ (x,p)\right\} : \nonumber \\
& &\begin{cases}
p=p_{2}-3(x_{2}-c)\ \mathrm{mod}(1)=x_{0}-x_{2}\ \mathrm{mod}(1), \\ 
x=x_{2}+p\ \mathrm{mod}(1)=x_{0}\ \mathrm{mod}(1).
\end{cases}
\label{px3}
\end{eqnarray} 
The result (\ref{px3}) and Eq. (\ref{B}) imply Eq. (\ref{mb2}).

\begin{center}
\textbf{APPENDIX C: AIs AND THE SHAPE OF} $\boldsymbol{C}$\\[0pt]
\end{center}

We study here the shape of the region $C$ for $K=3$\ in several cases. Let us
write Eq. (\ref{Cs}) as $C=C^{\prime }=B\cup \bar{B}^{(1)}\cup B^{(2)}$,
i.e., the union of the three sets in Fig. 10. This union is shown in Fig.
11, exhibiting a case in which both the $L$ and $R$ accelerator-mode islands
(AIs) exist (see Secs. IIIC and IV).
\begin{figure}[th]
\centering\includegraphics[width=7.0cm,height=7.3cm]{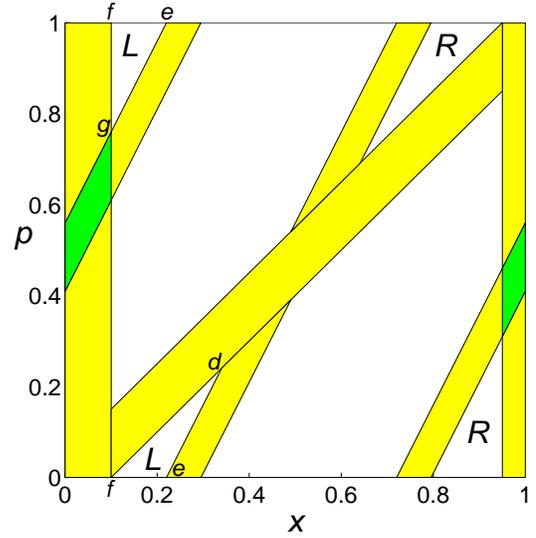}
\caption{(Color online) The region $C$, given by the union of the three
regions shown in Fig. 10, with overlaps indicated by green (dark grey) color. 
Both AIs, $L$ and $R$, exist in this case. See text for more details.}
\end{figure}
For values of $b_{1}$ and/or $b_{2}$ larger than those in Figs. 10 and 11,
there may exist only one AI or no AIs (see Fig. 3). A case of $C$ for which
only the $R$ AI exists is shown in Fig. 12 and is clearly different from
that in Fig. 11.
\begin{figure}[th]
\centering\includegraphics[width=7.0cm,height=7.3cm]{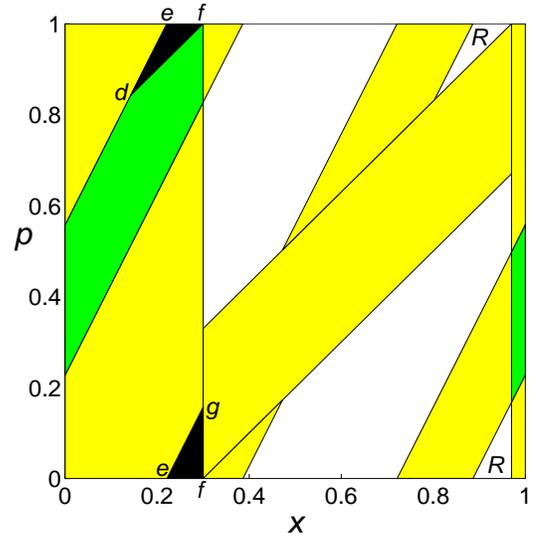}
\caption{(Color online) The region $C=B\cup \bar{B}^{(1)}\cup B^{(2)}$ for $K=3$, $b_1=0.3$, and $b_2=0.03$, with overlaps indicated by green (dark grey) color. Only the $R$ AI exists in this case. See text for more details.}
\end{figure}
We show below that the existence of AIs and the shape of $C$ in different cases depends on the location of the vertices $(x^{(j)},p^{(j)})$ ($j=1,2,3,4$) of the parallelogram (\ref{B2e}), shown in Fig. 10(c), relative to the strip $B$. Because of Eq. (\ref{B2}), one has $(x^{(j)},p^{(j)})=$ $\bar{M}(\bar{x}^{(j)},\bar{p}^{(j)})$, where $(\bar{x}^{(j)},\bar{p}^{(j)})$ ($j=1,2,3,4$) are the vertices of the parallelogram (\ref{B1e}) in Fig. 10(b). Clearly,
\begin{eqnarray}
(\bar{x}^{(1)},\bar{p}^{(1)}) & = & (b_{1},0),\ \ \ \ \ \ (\bar{x}^{(2)},\bar{p}
^{(2)}) =(b_{1},b), \label{xpb} \\
(\bar{x}^{(3)},\bar{p}^{(3)}) & = & (1-b_{2},1),\ (\bar{x}^{(4)},\bar{p}
^{(4)}) =(1-b_{2},1-b), \nonumber
\end{eqnarray}
where $b=b_{1}+b_{2}$. To derive explicit expressions for 
$(x^{(j)},p^{(j)})=$ $\bar{M}(\bar{x}^{(j)},\bar{p}^{(j)})$, one has to
properly determine the additive integers from the modulo operations in 
$\bar{M}$ so that $(x^{(j)},p^{(j)})$ will lie within the basic torus $\mathbb{T}^{2}$. We find that the values of $x^{(j)}$ are
\begin{eqnarray}
& &x^{(1)}=3c-2b_{1}-1,\ x^{(2)}=x^{(4)}=x^{(1)}+b, \nonumber \\
& &x^{(3)}=x^{(1)}+2b,   \label{xj}
\end{eqnarray}
indeed satisfying $0<x^{(j)}<1$ in the relevant cases in which at least one AI exists. In fact, in these cases one has $b<0.5$ [from Eq. (\ref{ccs}) with $K=3$] and the latter inequality implies by simple algebra that the smallest value of $x^{(j)}$ in Eqs. (\ref{xj}), i.e., $x^{(1)}$, satisfies $x^{(1)}>0$ while the largest
value ($x^{(3)}$) satisfies $x^{(3)}<1$. In addition, it is clear from Figs.
10-12 that the $R$ AI exists only if $x^{(3)}<1-b_{2}$ (vertex 3 is outside $B$); it is easy to show that the latter inequality is indeed equivalent to the existence
condition $b_{1}<F(b_{2})$ for the $R$ AI, derived in Sec. IIIC. Similarly,
the $L$ AI exists only if $x^{(1)}>b_{1}$ (vertex 1 is outside $B$), which can be easily shown to be equivalent to the existence condition (\ref{lai}). Thus, when both AIs exist, $b_{1}<x^{(j)}<1-b_{2}$ ($j=1,2,3,4$).

To determine the values of $p^{(j)}$, we first notice that the vertices 
$(x^{(j)},p^{(j)})$ must touch the boundaries of the region (\ref{B1}); 
this is because the vertices (\ref{xpb}) in Fig. 10(b) obviously touch 
the boundaries of the strip $B$ in Fig. 10(a) and 
$(x^{(j)},p^{(j)})=$ $\bar{M}(\bar{x}^{(j)},\bar{p}^{(j)})$. Then, in the
case that both AIs exist, i.e., $b_{1}<x^{(j)}<1-b_{2}$ (see above), 
$(x^{(j)},p^{(j)})$ touch the boundaries $p=x-b_{1}$ and $p=x+b_{2}$ of the
parallelogram (\ref{B1e}) [see Figs. 10(b), 10(c), and 11], so that
\begin{equation}
p^{(1,2)}=x^{(1,2)}-b_{1},\ \ \ p^{(3,4)}=x^{(3,4)}+b_{2}.  \label{pj}
\end{equation}
Assume now that only the $R$ AI exists, as in Fig. 12. Then, $x^{(1)}\leq
b_{1}$ (from above), i.e., vertex 1 (the point $d$ in Figs. 11 and 12) lies 
within the left part of strip $B$, on the boundary of region (\ref{B1}) given by 
$p=x-b_{1}\ \mathrm{mod}(1)$; thus, for $x^{(1)}<b_{1}$ (as in Fig. 12), $p^{(1)}$ in Eq. (\ref{pj}) must be replaced by $x^{(1)}-b_{1}+1$ while $p^{(j)}$ for $j>1$
remains unchanged. Similarly, when only the $L$ AI exists, vertex 3
lies within the right part of strip $B$, on the boundary of region (\ref{B1}) 
given by $p=x+b_{2}\ \mathrm{mod}(1)$; for $x^{(3)}>1-b_{2}$, $p^{(3)}$ in Eq. (\ref{pj}) must be replaced by $x^{(3)}+b_{2}-1$.

\begin{center}
\textbf{APPENDIX D: AREAS OF AIs}\\[0pt]
\end{center}

Consider the $L$ AI in Fig. 11. This is the triangle $dfg$ on the torus 
$\mathbb{T}^{2}$, composed of two triangles, $def$ and $efg$. The point $d$ is vertex 1 in Fig. 10(c) and the segment $de$ is part of the upper boundary of the region (\ref{B2e}). This boundary is a line of slope $2$ passing through vertex 1:
\begin{equation}
p-p^{(1)}=2(x-x^{(1)}).  \label{px1}
\end{equation}
Then, since $p_{e}=0$, we get from Eqs. (\ref{xj})-(\ref{px1}) that $x_{e}=(3c-b_{1}-1)/2$. Also, $x_{f}=b_{1}$ and $p_{f}=0$. The
point $g$, with $x_{g}=b_{1}$, lies on the line (\ref{px1}) with $p^{(1)}$
replaced by $p^{(1)}+1$. Thus, $p_{g}=2+3b_{1}-3c$. The area of the $L$ AI
is therefore
\begin{eqnarray}
S_{L} &=&S_{def}+S_{efg}=\frac{1}{2}(x_{e}-x_{f})[p^{(1)}+(1-p_{g})]  
\nonumber
\\
&=&\frac{1}{2}(3c-3b_{1}-1)^{2}.  \label{SL}
\end{eqnarray}
By symmetry arguments, the area of the $R$ AI is obtained from Eq. (\ref{SL}) by inserting the expression for $c$ from Eqs. (\ref{mmc}) and performing
the exchange $b_{1}\leftrightarrow b_{2}$. We get
\begin{equation}
S_{R}=\frac{1}{2}(2-3b_{2}-3c)^{2}.  \label{SR}
\end{equation}

\begin{center}
\textbf{APPENDIX E: AREA OF} $\boldsymbol{C}$\\[0pt]
\end{center}

The area of $C$ can be calculated starting from the relation $C=B\cup \bar{B}
^{(1)}\cup B^{(2)}$ (see above), where $\bar{B}^{(1)}$ and $B$ do not
overlap. Then, because of Eqs. (\ref{Bb1}) and (\ref{B2}), also $B^{(2)}$ does not overlap with $\bar{B}^{(1)}$. However, it may overlap with $B$. The area of $C$ is thus given by
\begin{equation}
S_{C}=S_{B}+S_{\bar{B}^{(1)}}+S_{B^{(2)}}-S_{B\cap B^{(2)}}.  \label{SC}
\end{equation}
From Eq. (\ref{B}), $S_{B}=b$, where $b=b_{1}+b_{2}$. The region 
$\bar{B}^{(1)}$ in Eq. (\ref{B1e}) is a parallelogram with basis $b$ (in
the $p$ direction) and height $1-b$ (in the $x$ direction), see also
Figs. 11 and 12. Then, $S_{\bar{B}^{(1)}}=b(1-b)$. From Eq. (\ref{B2}) 
and the fact that $\bar{M}$ is area preserving, it follows that 
$S_{B^{(2)}}=S_{\bar{B}^{(1)}}$. Finally, concerning the overlap $B\cap
B^{(2)}$, we consider first the case that both AIs exist, see Fig. 11. In
this case, $B\cap B^{(2)}$ consists of the green (dark grey) regions in Fig. 11.
These are two parallelograms having heights $b_{1}$ and $b_{2}$ (in the $x$
direction) and basis $b$, i.e., the width of region (\ref{B2e}) in the 
$p$ direction. Thus, $S_{B\cap B^{(2)}}=b^{2}$. The area (\ref{SC}) is
therefore
\begin{equation}
S_{C}=3(b_{1}+b_{2})-3(b_{1}+b_{2})^{2}.  \label{SC1}
\end{equation}
Consider now the case that only one AI exists, say the $R$ AI as in Fig.
12. In this case, as explained at the end of Appendix C, the point $d$, i.e., the vertex 1 of region $B^{(2)}$, lies inside the left part of strip $B$, on the boundary of region (\ref{B1}). This means that the black triangles $def$ and $efg$
in Fig. 12 are not included in the region $B^{(2)}$ or $B\cap B^{(2)}$ but
they are actually part of the region (\ref{B1}). Thus, to calculate $S_{B\cap B^{(2)}}$ one must subtract from $b^{2}$ (the value of $S_{B\cap B^{(2)}}$ in the previous case) the areas of $def$ and $efg$. The area (\ref{SC}) is then obtained by adding $S_{def}+S_{efg}$ to the expression (\ref{SC1}). By comparing Fig. 12 with Fig. 11, it is clear that the areas $S_{def}$ and $S_{efg}$ can be calculated precisely as in Appendix D and $S_{def}+S_{efg}$ is given again by formula (\ref{SL}). Therefore, the area (\ref{SC1}) increases precisely by an amount equal to the area (\ref{SL}) of the missing $L$ AI:
\begin{equation}
S_{C}=3(b_{1}+b_{2})-3(b_{1}+b_{2})^{2}+(3c-3b_{1}-1)^{2}/2.  \label{SC2}
\end{equation}
Similarly, when only the $L$ AI exists, one must add the area (\ref{SR}) to 
(\ref{SC1}):
\begin{equation}
S_{C}=3(b_{1}+b_{2})-3(b_{1}+b_{2})^{2}+(2-3b_{2}-3c)^{2}/2.  \label{SC3}
\end{equation}


\begin{thebibliography}{99}
\bibitem{note} One can also easily show that in condition (\ref{lai}) (with $2\leq K<4$) one must have $b_1<b_1^{(-)}$, where $b_1^{(-)}=[2K-1-\sqrt{4K+1}]/(2K)$ is the smallest root of $F(b_1)=0$ and $F(b_1)$ decreases monotonically from $F(b_1=0)=(K-2)/(K-1)$ to $F(b_1=b_1^{(-)})=0$. In addition, $b_1^{(-)}<b_1^{(0)}$, where $b_1^{(0)}=(K-1)/K$ is the value of $b_1$ at which the denominator of $F(b_1)$ vanishes. Thus, this denominator is always positive in condition (\ref{lai}). Similar results hold for analogous conditions in Sec. IIIC.
\end{thebibliography}
\end{document}